\documentclass[12pt,aps,prd,preprint,showkeys]{revtex4}
\usepackage[utf8]{inputenc}
\usepackage{amsmath}
\usepackage{amsfonts}
\usepackage{amssymb}
\usepackage{graphicx}
\usepackage{csquotes}
\RequirePackage[T1]{fontenc}
\usepackage{url}

\begin{document}
\title{Gamma Ray Bursts from magnetic reconnection in the cosmic string wakes}
\author{Dilip Kumar \thanks{21phph03@uohyd.ac.in}}
\author{Bijita Bose \thanks{22phph10@uohyd.ac.in}} 
\author{Soma Sanyal \thanks{somasanyal@uohyd.ac.in} }
\affiliation{School of Physics, University of Hyderabad, Central University P.O, Hyderabad-500046, Telangana, India.}


\begin{abstract}
Magnetic reconnection in magnetized wakes of cosmic strings results in the release of a large amount of energy. This energy is released in a short period of time. In this work, we show that this sudden release of energy can result in a Gamma Ray Burst (GRB) of short duration. The magnetic reconnection occurs at several points of the cosmic string wake. The emerging shocks from these points have different velocities. These shocks will collide with each other and give rise to short bursts of energy. The emitted pulse of energy depends on the background magnetic field and the timescale associated with the magnetic reconnection in the cosmic string wake. We also obtain the synthetic lightcurve that occurs from multiple collisions of the shock waves that are emitted from the multiple points of magnetic reconnection in the wake region. Finally, we fit the current experimental data of short GRB which are in the same energy range as the ones predicted from our model.  

\end{abstract}

\maketitle            
\section{Introduction}								\label{sec:Intro}

Symmetry breaking phase transitions in the early universe give rise to topological defects. One of the topological defects that is predicted by these symmetry breaking phase transitions is the cosmic string defect. The cosmic string defect is a one dimensional defect generated due to a $U(1)$ symmetry breaking phase transition.  Once formed, these strings will move through the early universe plasma. Depending on the exact nature of the phase transition, long strings or loops of cosmic strings are formed. Some of these defects are unstable and decay with the evolution of the universe, others are stable and evolve as a giant network of cosmic strings spanning the universe \cite{cosmicstrings}. Theoreticians have predicted multiple observational signatures of these cosmic strings. Primary amongst these  were the gravitational lensing of objects due to the conical space time of the cosmic string \cite{lensing}. Other than the gravitational lensing, as the cosmic string moves through the plasma, it gives rise to wakes of overdensity behind it \cite{wakes}. These wakes will lead to a distinct wedge shaped patterns in the $21$ cm redshift survey maps \cite{branden1,dacunha}. These wakes also give rise to temperature fluctuations which have been used for searches for these exotic objects by the Planck collaboration \cite{Planck_2016}. Constraints on the deficit angle of the cosmic string has been studied from the radio maps of CMB anisotropy, provided by the space mission Planck for various frequencies \cite{Sazhina_2014}.  All these observational signatures are currently from density inhomogeneities generated as the cosmic string moves through the plasma. Recently, there have been some signatures of Gravitational waves generated by these cosmic string networks \cite{Blasi_2021}.   

Cosmic strings moving through the plasma have also been associated with the primordial magnetic fields in the early universe. Two cosmic strings moving past one another in the plasma lead to the generation of primordial magnetic field in the early universe \cite{vachaspati}. The Harrison mechanism which leads to differential rotation between the heavier ions and the lighter electrons also leads to the generation of a primordial magnetic field close to the cosmic strings \cite{Harrison}. In a recent work, we have shown that magnetic fields can be generated in cosmic string wakes due to the Biermann battery mechanism \cite{Sovan_2020}. 

The wakes of these cosmic strings are long and narrow. The Biermann battery mechanism \cite{ohira} produces magnetic fields that are oppositely directed along the two sides of the wake. In the next section, we will give a brief discussion on the generation of the oppositely directed magnetic fields in the cosmic string wake and the neutral line between the fields where the points of magnetic reconnection occur. Detailed discussion on these phenomena can be found in our previous studies reported in ref.\cite{Sovan_2020} and ref.\cite{Dilip_2023}.

Magnetic reconnection occurs when two oppositely directed magnetic fields come very close to each other. It leads to the rearrangement of the magnetic field lines by breaking and cross-connecting to each other which results in a burst of energy. Numerical simulations using magnetohydrodynamic codes have also shown the evolution of the magnetic field and the formation of loops of magnetic field lines due to magnetic reconnection \cite{Soumen_2022} in cosmic string wakes. We had in our previous work predicted that this magnetic reconnection can lead to a Gamma Ray Burst (GRB) \cite{Dilip_2023}. GRB's are the bursts of energies observed by various telescopes. Their extreme luminosity and high energy output make them crucial for studying the early universe \cite{Paczynski_1998}.

Previously GRBs have been associated with superconducting cosmic strings \cite{Babul_1987}, where the electromagnetic radiation is powered by the electric currents flowing through the string cores. Cusps of superconducting strings also emit powerful beamed pulse of electromagnetic radiation \cite{berezinsky} which can be identified with GRBs. The only possibility of GRBs from non-superconducting cosmic strings was our prediction of a GRB from magnetic reconnection in the magnetized wake of a cosmic string \cite{Dilip_2023}. 
Magnetic reconnection has been considered a progenitor for a GRB in many other scenarios \cite{Granot}. Depending on the different models of magnetic reconnection, shocks are usually generated at the points of reconnection \cite{yamada_review}. In our previous work, we did not consider the details of the magnetic reconnection in cosmic string wakes, therefore we could give only the total energy or fluence of the GRB that is generated. In this work, we look at the points of magnetic reconnection in cosmic string wakes in greater detail and obtain a detailed light curve of a possible GRB from the wake of a magnetized cosmic string. In the summary, we will also discuss how the GRBs from the cusps of superconducting strings will be different from the ones in the wake of the cosmic string.   

We have shown in a previous work which we discuss briefly in the next section that magnetic reconnection can occur in the wakes of cosmic strings. Due to the structure of the wake,  multiple points of magnetic reconnection will occur. As the magnetic lines rearrange themselves, shocks can occur in the outflow. Slow mode shocks have been predicted in the Petschek model \cite{Petschek} and fast mode shocks have been predicted for models with fast reconnection.  In a cosmic string wake, these reconnection points are close to each other and therefore the shocks emanating from the points of magnetic reconnection will collide with one another and merge to form newer shocks. The merging of the shocks will lead to a sudden increase of the total energy in the plasma. This will cause the acceleration of the particles in the plasma. This sudden relativistic acceleration will lead to the radiation of non-thermal energy which can be identified as a GRB. We calculate the energy released during this process and obtain the shape of the pulse of a GRB light curve based on the parameters of the cosmic string wake. Since multiple such pulses will be generated, we also consider the case of large number of pulses with different velocities coming together to form the GRB light curve.    

The GRB's are generally classified based on their luminosity and the time duration over which they occur \cite{piran}. These are two of the basic features of any GRB. The short duration GRBs last for less than two seconds while the long duration GRBs usually last for more than two seconds. The GRB light curves we obtain are for high-redshifts and are of short-duration. We also find that the possible time durations and the fluence calculated in our model is within the range of the high-redshift, short duration GRBs identified by the satellites like FERMI and SWIFT. Overall, we establish that magnetic reconnection in cosmic string wakes may provide a new way of identifying these exotic objects using both electromagnetic signals and signatures from  density inhomogeneities.

\section{Magnetic Reconnection in Cosmic String Wakes} 			\label{sec:MR}

In a recent work, we have shown that the small scale inhomogeneities which are perpendicular to the shock fronts in the cosmic string wakes generate  magnetic fields in the wake due to the Biermann battery mechanism \cite{Sovan_2020}. We will now briefly describe this scenario in this section. We will then proceed to discuss how mulitple points of magnetic reconnection can occur in the cosmic string wake in the next subsection.

\subsection{Biermann battery mechanism in the wakes of cosmic strings}

We know from the literature, that at any given time $t_i$, the cosmic string wake has dimensions given by \cite{branden1}, $ c_{1} t_{i} \times t_{i} v_{s} \gamma_{s} \times \delta \theta t_{i} v_{s} \gamma_{s}$ where, $c_{1}$ is of order $1$, $v_{s}$ is velocity of the moving string, $\gamma_{s}$ is the Lorentz factor associated with the velocity of the string. Since, the string usually moves at relativistic velocities, the average velocity is given by $<v_{s}\gamma_{s}>$ $\sim 0.5$ \cite{dacunha}. The $\delta \theta$ in the dimensions is the opening angle of the cosmic string wake. It is equal to the deficit angle of the space time around the cosmic string and is given by $\delta \theta \sim 4 \pi G\mu$. The value of $G\mu$ is given by the symmetry breaking scale at which the cosmic string is formed \cite{vilenkin}. Since this is  small (of the order of $10^{-7}$ or less), thus the width of the wake is  small compared to the length of the wake.  

The cosmic string wakes are thus narrow with the length of the order $\sim 10^{28}$ cm at around redshift $z \sim 6.7$ and a width of the order $\sim 10^{22}$ cm. The wake itself has an overdensity of a height approximately given by $t_i v_s \gamma_s$, where $t_i$ is the time at which the wake is being observed and $v_s \gamma_s$ is the relativistic speed at which the cosmic string is moving through the plasma. The plasma itself is not stationary and the plasma particles are dragged into the wake of the string due to the conical space-time of the cosmic string metric. So, both the plasma particles as well as the cosmic string are moving and the velocity we talk about is the relative velocity of the particle in the reference frame of the cosmic string. Due to this large overdensity of the wake, a temperature gradient is formed across the wake \cite{ssanyal1,ssanyal2}. On the two sides of the wake we have a lower temperature and the wake itself is at a higher temperature. So there are two oppositely directed temperature gradients on both sides of the wake. The wake itself is not homogeneous with significant inhomogeneities being generated due to the different masses of the particles in the wake. We have shown in ref. \cite{Sovan_2020} that a neutrino current is generated close to the cosmic string which is oscillatory in nature. This oscillatory neutrino current drives the electrons to generate a correponding oscillating electron 
current. Thus, electrons and positively charged particles are distributed inhomogeneously in the wake of the cosmic string. The gradient of the density inhomogeneity is roughly perpendicular to temperature gradient on the two sides of the wake, so, a magnetic field is generated in the wake. It has also been shown previously that particles of different masses cluster at different distances from the cosmic string \cite{sornborger}, this again leads to density inhomogeneities in the cosmic string wake. These inhomogeneities are along the wake and therefore perpendicular to the temperature gradient. Thus according to the Biermann mechanism, they will generate magnetic fields as the term $\nabla \rho \times \nabla T$ will be non-zero (i.e. the density gradients are not aligned with the temperature gradient).  

\subsection{Magnetic reconnection in the wakes} 

As mentioned before the temperature gradient is oppositely directed in the cosmic string wake [For figure: refer to \cite{Sovan_2020}]. So due to the  symmetry of the wake, the fields generated are oppositely directed on the two sides of the wake. 
\begin{figure}[h]
	\centering
	\includegraphics[width=0.5\textwidth]{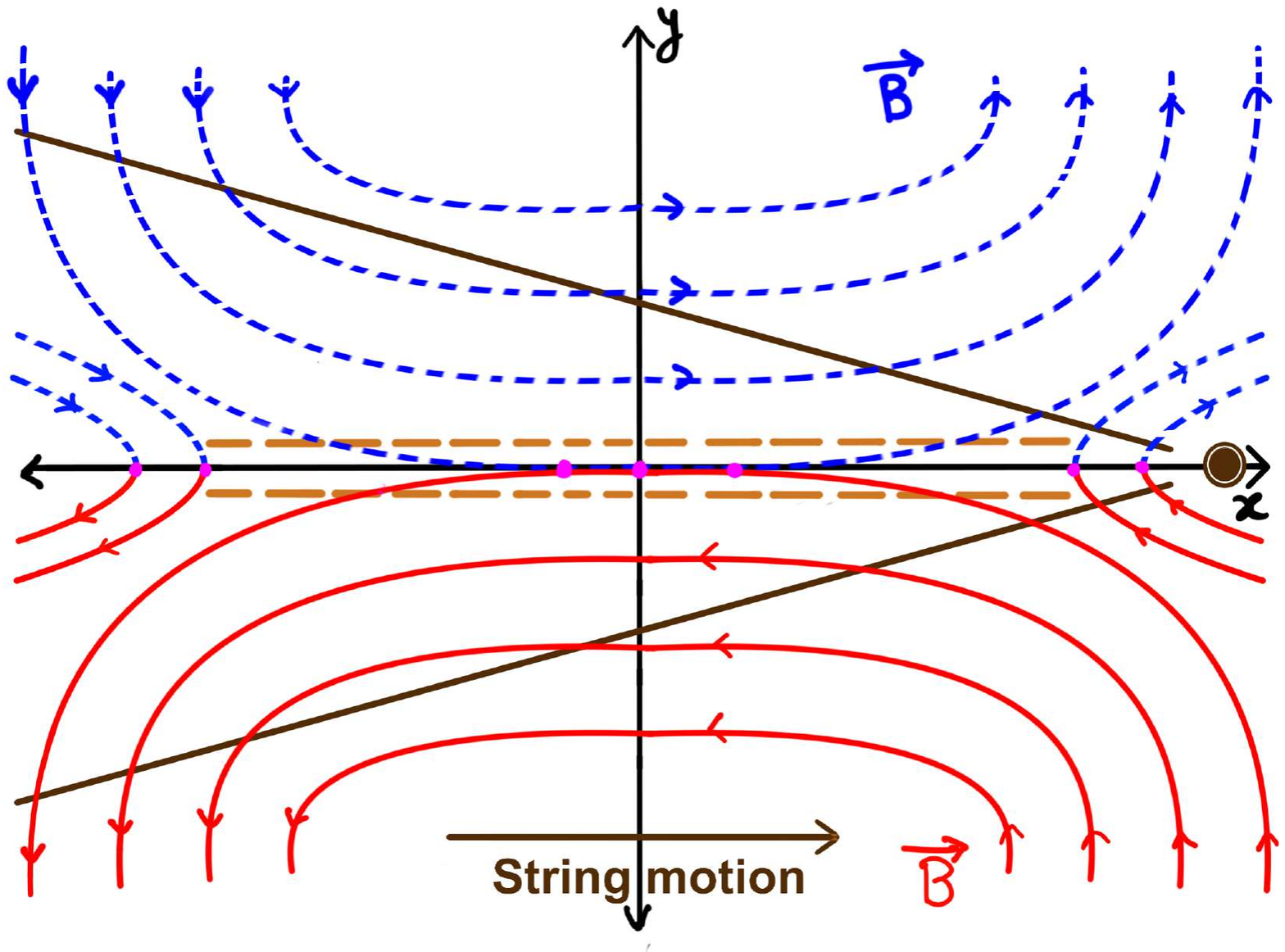}
	\caption{The schematic diagram shows a cosmic string (marked by $\odot$) on the $X$-axis, moving through the plasma. The cosmic string is perpendicular to the plane of the paper. As it moves, a wake forms behind the string, outlined by solid brown lines. The opposite magnetic fields are then generated in this wake by the Biermann battery mechanism, with their directions shown by blue dashed and red solid lines. Points of magnetic reconnection being formed in the cosmic string wake shown by small dots (magenta).}
	\label{fig:reconnection}
\end{figure}
In Fig.\ref{fig:reconnection}, we have shown that the cosmic string is moving to the right while the plasma flows to the left. The two dark brown (solid) lines depict the two sides of the cosmic string wake which is symmetric about the $X-$axis of the coordinate system considered. Since we are only looking at the $X-Y$ plane, the magnetic field lines (solid red line and dashed blue line) are shown on both sides of the $X-$axis. As seen from the figure, the temperature gradient is oppositely directed on two sides of the wakes along the $Y-$axis. The density inhomogeneities are along the $X-$axis. So the magnetic fields are along the positive and negative $Z-$direction (not on the depicted plane). Thus, the magnetic field from Biermann battery mechanism is given by, 
\begin{equation}
	\vec{B} ~ = ~ -\hat{i} \frac{\partial N_e}{\partial z} \frac{\partial T}{\partial y} ~+~ \hat{k} \frac{\partial N_e}{\partial x} \frac{\partial T}{\partial y}~ 
\end{equation}
The width of the wake overdensity is small as mentioned previously so a quasi two dimensional study 
is justified so the subsequent field is approximated by, $\vec{B} = -\hat{i} \frac{\partial N_e}
{\partial z} \frac{\partial T}{\partial y} $ for the quasi- 2 dimensional picture. The diffusion 
region is marked with the dashed line parallel to the $X-$axis and small dots (magenta color) are 
put to indicate the possible points of reconnection in Fig.\ref{fig:reconnection}.

Generally due to the frozen in condition of the magnetic field, the magnetic field lines are deformed by the plasma flow. So as the background particles in the plasma stream into the wake overdensity with relativistic velocities, they push the magnetic field lines together and magnetic reconnection occurs. The oppositely directed magnetic fields are along the $Z-$direction so the relevant length would be along the $Z-$direction. As mentioned, a quasi 2-dimensional model is used to understand magnetic reconnection,  the dimensions of the cosmic string wake indicate that the two oppositely directed magnetic field are separated by a very small lengthscale. So a quasi two dimensional model can be used to understand the magnetic reconnection in the cosmic string wake.

A basic calculation based on the Sweet–Parker (SP) model \cite{Parker_1957} shows that a large amount of energy is released due to the magnetic reconnection in the cosmic string wake. The SP model approximated the problem as a quasi-two dimensional problem with a plasma of constant resistivity. The energy though large was released over a substantial period of time. In the current work, we do not use the SP model. Instead we will work with faster reconnection models such as the Petschek model. There are other models which increase the reconnection rate by using a variable resistivity \cite{che}.

The Lundquist number is very high for these astrophysical plasmas. The typical length of the wake is large and the resistivity is low, so the Lundquist number for our case is obtained of the order of $\sim 10^{22}$. Such high numbers have previously been found for different scenarios of astrophysical conditions \cite{lundquistno}. At such high Lundquist numbers, fast magnetic reconnection occurs often leading to the formation of mulitple plasmoids.   
\begin{figure}[h]
	\centering
	\includegraphics[width=0.5\textwidth]{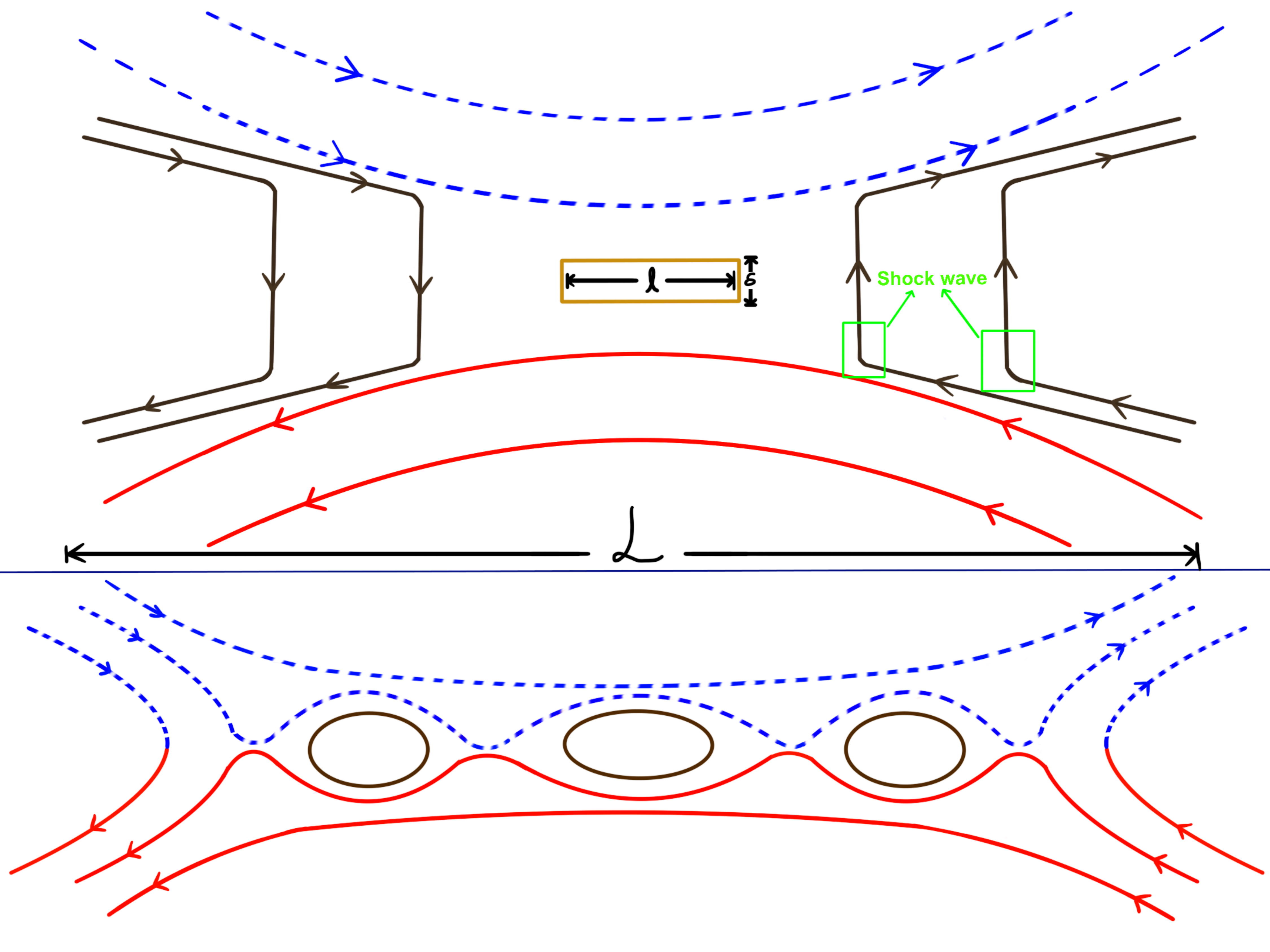}
	\caption{The Schematic figure represents the regions of fast magnetic reconnection and plasmoids formation. The upper part of the figure shows the generation of shocks due to fast reconnection, while the lower part depicts the multiple reconnection events that lead to the formation of plasmoids. The two regions are presented separately for clarity.}
	\label{fig:plasmoids}
\end{figure}
In Fig.\ref{fig:plasmoids}, we have shown the region of fast magnetic reconnection in the upper part of the figure and the mulitple plasmoids being formed in the lower part of the figure. The two parts are separate. The upper one depicts the formation of shocks due to fast magnetic reconnection, while the one below shows that mulitple reconnection which form the magnetic islands or plasmoids. In a previous numerical simulation of magnetohydrodynamics in a cosmic string wake, mulitple such islands have been seen. [See for ref.\cite{Soumen_2022} for details of the numerical simulations of magnetic reconnection in cosmic string wakes.]

\section{Shocks generated in the wake} 

Hydrodynamic shocks have previously been studied in detail in the wakes of cosmic strings \cite{sornborger}. The shocks are formed as the cosmic string move with relativistic speed through the early universe plasma. The particles in the plasma streaming past the moving string generate the wake as shown in Fig.\ref{fig:reconnection}. The dark brown solid lines in Fig.\ref{fig:reconnection} denote the boundaries of the wakes where hydrodynamic shocks are typically generated. However, shocks in a magnetized wake has not been studied in great detail. Some recent studies show the formation of shocks in a magnetized wakes of cosmic string \cite{Soumen_2022}. Both weak as well as strong shocks can be formed in the wake depending on the relative velocity of the plasma particles and the cosmic string. Alfven's theorem ensures the squeezing of the magnetic field lines in the cosmic string wake due to the inward streaming of the plasma particles but Alfven's theorem breaks down at small lengthscales and multiple X-points of magnetic reconnection are obtained in the width of the wake \cite{Bist2025}.

It is well known in the literature that shocks are generated in the outflow of the Petschek and other fast reconnection models \cite{yamada_review}. These shocks will move outward and on coming in contact to large density inhomogeneities will be reflected as stronger shocks. So within the cosmic string wake we will have shock-shock collisions. 

Usually the shocks generated in the Petschek model are the slow mode shocks. But faster shocks can also be generated due to other models of reconnection. There are studies of fast reconnection in astrophysical plasmas. One of the context is when magnetic reconnection occurs in turbulent MHD \cite{Eyink}. In such cases, the magnetic islands or plasmoids may not be aligned with the symmetry axis of the cosmic string wakes as shown in Fig.\ref{fig:plasmoids}. Thus in the literature, there is a variety of plasma reconnection scenarios which give rise to shock structures \cite{Workman} and it is possible to generate fast mode shocks in magnetic reconnection outflows. Even if there are weak shocks generated in the wake of the cosmic string, they will collide with the shock from the wake boundary and be reflected back as stronger shocks. Therefore, there will be shock-shock collisions in the cosmic string wake structure. There are many different scenarios in which the collision can take place. In this case, we do the calculation for one such scenario. There are other different possibilities that may arise but we believe the results would be generically similar to the current case. In the next section, we discuss the model of Shock -Shock collision we follow and give the detailed equations.

\section{Models of Shock-Shock collisions} \label{sec:shockcollisions}

There are many models of shock-shock collision which have been studied both analytically as well as numerically \cite{Hartigan,Rahaman}. There have also been models of collision of plasmoid blobs \cite{meng,yuan}, However we are considering the shocks as shells emanating from the reconnection points with variable Lorentz factors. One of the models that has been used in such a scenario is the hydrodynamic shell collision model \cite{Asaf}. Here, the shocks are typically modeled as hydrodynamic shells of different densities and different velocities. They may be initiated from the same point or different points. Similarly the collision between the shells may be head on or may be at an oblique angle. Shocks may merge together after collision or they may be reflected by the other shock front. Here we work on only one scenario out of all these. We consider a model where a slow shock is followed by a faster shock. The energy reaches it's peak as the two shocks merge together and move with the combined total kinetic energy. The particles of the plasma are accelerated to extremely high velocities at the moment of the collision and then decelerated as the merged shock gradually slows down. The accelerated charged particles moving in the magnetized wake of the cosmic string will then release a considerable amount of non-thermal radiation in a short time which can be identified as a GRB.  
 
We consider the two shocks to be emanating from the same point one after the other after a  small interval of time. Both the shocks move with relativistic velocities but the first shock is assumed to be slower than the second shock. We have considered that the slower shock, has a density of $\rho_1$ and a relativistic velocity given by $\gamma_1 v_1$. The second shock is the faster shock with a density of $\rho_2$, having a relativistic velocity $\gamma_2 v_2$. The kinetic energy density of individual shocks from magnetic reconnection is given by \cite{Dilip_2023},  
\begin{equation}
(K.E.)_{i} = \dfrac{1}{2} \rho_i \gamma_{i}^2 c^{2} \frac{B}{(4 \pi \rho)^{1/2} \sigma l}. 
\end{equation}
The subscript $i$ will be $1$ or $2$ depending on the shock being considered and $\rho$ is the density of the plasma. After the collision, the two shocks will merge together and move as a single shock. So, after collision we denote the merged density as $\rho_{m}$ and the merged Lorentz factor as $\gamma_{m}$. For different models of reconnection, the length of the diffusion region $l$ is what makes a difference in the rate of reconnection and the energy emitted \cite{Kulsrud, yamada_review}. In the fast reconnection models, the length is scaled due to the emergent shocks which often increases the overall reconnection rate. In the Petschek model \cite{Petschek}, the length of the diffusion region $l$ is smaller than the diffusion region of Sweet-Parker model, and the width of diffusion region can be calculated from this length. The width is further scaled by the ratio of the magnetic fields. 

Here we are considering that the reconnection occurs in the wake of the cosmic string. So the cosmic string parameter $G\mu$ will determine the length and the width of the reconnection region. We have considered the energy scale for the symmetry-breaking phase transition for Abelian Higgs string to be of the order of $ 2.2 \times 10^{15}$ GeV \cite{Lizarraga_2016}. The order of $G\mu$ for such Abelian Higgs cosmic string is therefore approx $10^{-11} - 10^{-8}$.  There is also  the current observational limit from the Planck Collaboration \cite{Planck_2016} which gives the upper limit of the value of $G\mu$ as $\sim 10^{-7}$. We can use this to estimate the length of the reconnection region. The total length of the cosmic string wake is taken to be $L$. This is the longest dimension of the wake. Reconnection however does not occur for the full length of the wake. To estimate reconnection length $l$, we have to obtain the Alfven velocity. The Alfven velocity is given by $V_A = \frac{B}{\sqrt{\mu \rho}}$, where $\mu$ is the permeability and $\rho$ is the density of the plasma and $B$ is the magnetic field in the wake. Thus, the length of the reconnection region $l$ is obtained as, 
\begin{equation}
 l = \dfrac{c L}{(V_{A} \sigma L)^{1/2}},
\end{equation}
where, $\sigma$ is the conductivity of the plasma. We will henceforth consider $c = 1$ in all our calculations.

To understand how the kinetic energy changes with time, we will use a simple model of shock-shock collision where the mass of the shock changes with time. This change is however small and it occurs due to the continuous streaming of the plasma particles into the wake region. We first obtain the single pulse of energy produced due to the merger of two relativistic shocks with respect to the time interval it is generated in. This time interval is divided into three stages. The first stage is the situation where we have two individual shocks moving towards one another, before the collision, i.e. $t<t_{col}$ and the second stage is when the individual shocks collides and merger happens, i.e. $t=t_{col}$. This is the time at which the kinetic energy reaches it's peak. Subsequently, in the third stage the energy decreases as the merged shock spreads out, after the collision, i.e. $t>t_{col}$. The energy leads to the acceleration of the particle so the pulse generated in this time interval will determine the luminosity of the GRB emitted due to the radiation from the accelerated particles. One important point that we also tried to incorporate here is the decrease in the magnetic field in this stage. Downstream to the collision of the shocks it is known that the magnetic field decays \cite{Zhou_2023}. We have included this decreasing magnetic field in this stage. We now proceed to discuss each stage in some detail. 
  
The first shock is taken to be produced at $t = 0$ and the subsequent shock is produced after a time period $\Delta t$. So, the time duration of the collision will be $\Delta t = \dfrac{l_{r}}{2 \gamma_1^2 c}$ \cite{beniamani}. Here, $l_{r}$ denotes the total distance over which the collision takes place and the merger of the shocks happen. So, both the shocks travel the distance $l_r$ from the point at which they were generated at different velocities (and therefore different times $t_1$ and $t_2$). The kinetic energy of the first shock is given by $\dfrac{1}{2} m_1 \gamma_{1}^2 c^{2} \frac{B}{(4 \pi \rho)^{1/2} \sigma l}$ and the kinetic energy of the second shock by  $\dfrac{1}{2} m_2 \gamma_{2}^2 c^{2} \frac{B}{(4 \pi \rho)^{1/2} \sigma l} $. Before the merger, the total energy would be the  sum of these two energies. We are considering only two dimensional shocks here as the shocks are generated from a quasi two-dimensional model. The shocks are generated from the same point and have a radius $l$ at time $t$. As the shocks move with a velocity $v_s$, more particles are swept in continuously, so the mass of the shocks will be given on the area $l \times \delta$ by $m_{i} = \rho_i \times (l \times  \delta) $. The $\delta$ is the width of the reconnection region and is shown in Fig.\ref{fig:plasmoids}. The individual shock velocity $v_{is}^2 = (l \times \delta) /t^2 $. The $i$ in the subscript stand for the numbers $1$ or $2$ depending on which shock we are considering. Substituting all these in the kinetic energy equation we get the total energy as a function of time. So in the first stage at $t<t_{col}$, the total energy would be given by 
\begin{equation}
E_I(t)  = \dfrac{1}{2} \rho_1 v_{1s}^2 \gamma_{1}^2 c^{2} t^2 \frac{B}{(4 \pi \rho)^{1/2} \sigma l} + \dfrac{1}{2} \rho_2 v_{2s}^2 \gamma_{2}^2 c^{2} t^2 \frac{B}{(4 \pi \rho)^{1/2} \sigma l}.
\end{equation}
Here, the magnetic field $B$ is considered to be a constant. 
In the second stage, as the two shocks merge together, the energy reaches the maximum possible value. After this, the density of the combined shock wave is given by the density of the merged shocks $\rho_m$, the Gamma factor of the merged shocks $\gamma_m$ and the velocity of the shock at the merger point which is again denoted by the distance and the time. So, at the $t=t_{col}$, the distance at which the merger happens is $l_r$. Thus, the maximum energy is therefore given by, 
\begin{equation}
E_{II}(t)  = \dfrac{1}{2} \rho_m v_m^2 \gamma_{m}^2 c^{2} t_{col}^2 \frac{B}{(4 \pi \rho)^{1/2} \sigma l_{r}} = E_{max}. 
\end{equation}
Here, $\rho_m = (\rho_1 + \rho_2)/2$ and $v_m \gamma_m$ is given by, 
\begin{equation}
v_m \gamma_m = \sqrt{\frac{2 (\rho_{1} v_{1}^{2} \gamma_{1}^{2} + \rho_{2} v_{2}^{2} \gamma_{2}^{2} )}{\rho_{1} + \rho_{2}}} .
\end{equation}      
To simplify the calculations, we have considered the densities of the shocks to be the integral multiples (e.g, $\rho_2 = 2 \rho_1 $ etc). It has been shown that the collision of the shocks will also generate some random magnetic fields which will decay in a short time \cite{Zhou_2023}.  

Upto the point of collision, the total kinetic energy can increase but after that it will start to decrease as the merged shock gradually dissipates. In several shock collision models, where the particle acceleration has been studied using transport equations, the maximum distribution of the accelerated particles is found to be at the point of collision of the shocks \cite{vieu}. Since in this case we have not considered the particle distribution function, we denote the total energy of the shocks as the kinetic energy comprising of the mass density and the velocity of the shocks. 

Similar to ref.\cite{kobayashi}, we assume once the peak amplitude is obtained, the decrease in the energy is due to the angular spreading of the shock wave into the background plasma. Assuming the density will remain more or less the same, it will be the velocity which will determine the decrease in energy. As the timestep in the numerical model is small, a Taylor series expansion will result in the approximate increase of energy with time as $\Delta E \propto (1 - \frac{t_{col}^2}{t^2})$. The constant of proportionality will depend on the same parameters as the maximum energy.  So after collision, i.e. $t>t_{col}$, the energy will decrease as $E_{max} - \Delta E$. The kinetic energy expression after the collision in the third stage is given by, 
\begin{equation}
E_{III} (t) = E_{max} \left( 1 - (1- \frac{t_{col}^2}{t^2})  \right) .
\label{energy3}
\end{equation} 
Usually the kinetic energy leads to the radiation spectra which in most cases is modeled by the Band function. This function consists of a set of exponentially connected broken power laws. Since the magnetic field is an essential part of the kinetic energy dissipated,  we would like to model the decrease in the magnetic energy by a time dependent parameter. We consider this decreasing function using a function dependent on time and two other parameters which can be varied. The net result is that the magnetic field should be replaced by a polynomial decaying magnetic field as discussed in ref. \cite{Zhou_2023}. Since the expression for $E_{max}$ contains the magnetic field in equation 
\ref{energy3}. We replace the constant $B$ in this expression with a time dependent magnetic field. 
The field will decrease with time so at any given time the field will be proportional to $(1 -(t_{col}/t))$. Therefore the equation for the energy decay in the third time interval is replaced by
\begin{equation}
E_{III} (t) = \dfrac{1}{2} \rho_m v_m^2 \gamma_{m}^2 c^{2} t_{col}^2 \frac{B(t)}{(4 \pi \rho)^{1/2} \sigma l_{r}} \left( 1 - (1- \frac{t_{col}^2}{t^2})  \right) .
\end{equation}
Here, $B(t)$ denotes the time dependent magnetic field. 
Since we do not know the actual function, we introduce two variable parameters ($\alpha$ and
$\beta$) in the model. We will later on use these parameters to fit the experimental data. The expression therefore becomes 
\begin{equation}
E_{III} (t) = \dfrac{1}{2} \rho_m v_m^2 \gamma_{m}^2 c^{2} t_{col}^2 \frac{B}{(4 \pi \rho)^{1/2} \sigma l_{r}} \left( 1 - \left(1- \beta \left({\frac{t_{col}}{t}}\right)^{\alpha}\right)  \right) .
\end{equation}
As can be seen, $\alpha = 2$ and $\beta = 1$ gives us the case of the constant magnetic field. To 
study the different dependencies we have used these values. The parameters have been used later as 
free parameters to fit the experimental data.    
The reason we do this is because we do not know how the magnetic field will actually decay in our present scenario.

\begin{figure}[h]
	\centering
	\includegraphics[width=0.75\textwidth]{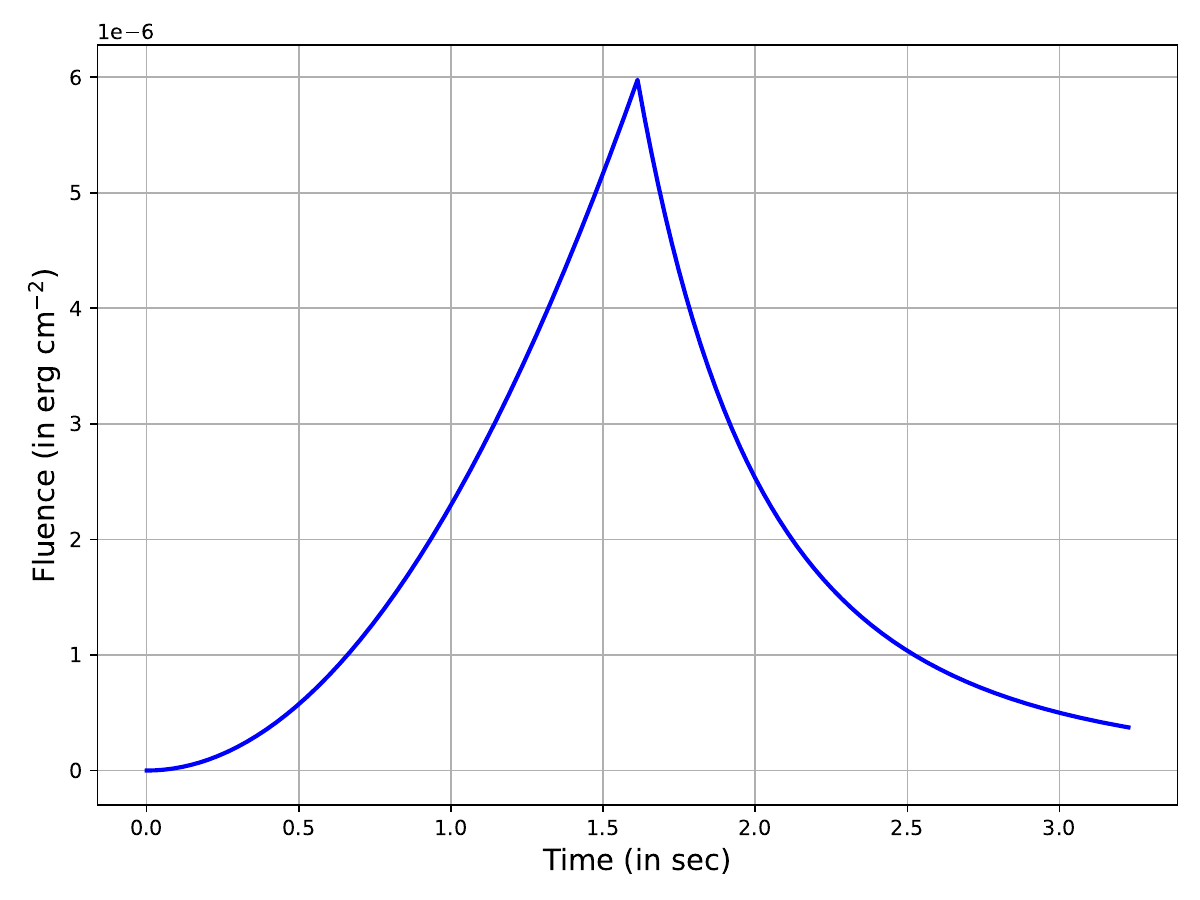}
	\caption{Change of energy fluence with time as two shocks waves collide and merge, followed by the decay.}
	\label{fig:lgpulse}
\end{figure}

Fig.[\ref{fig:lgpulse}] shows the resulting pulse that we have obtained from the above energy expressions for three different stages. 
We find that there is a sharp rise followed by a gentle decay. The collision which leads to the merger is considered instantaneous and hence we have obtained a sharp peak. In a more realistic scenario, it would probably happen over a short period of time which will lead to a more rounded peak. For this particular pulse, we have considered the time to be at redshift $z = 6.7$. We have calculated the background plasma density at that temperature and time. We have also taken the average magnetic field to be the background equipartition field at that time. This has a value of $1.95 \times 10^{-2} ~eV^2$. This is the minimum magnetic field at that temperature. Usually, the magnetic field is amplified at least by a factor of $10$ in the wake of the cosmic string. This happens due to the Alfven's theorem of flux freezing as there is an inward flow of streaming particles in the wake of the cosmic string. An enhanced magnetic field will give us GRB's in a higher energy band. We have estimated the electrical conductivity of the early universe plasma from the general expression \cite{Grasso_2001},
	\begin{equation}
	\sigma = \frac{n_{e} e^{2} \tau}{m_{e}},
\end{equation}
where $n_{e}$ is the electron number density, $e$ is the electron charge, $\tau$ is the mean collision time, and $m_{e}$ is the electron mass. For post-recombination plasma, the electron number density is given by,
\begin{equation}
	n_{e}(z) \sim 3 \times 10^{-10} cm^{-3} ~ \Omega_{0} h ~ (1 + z)^{3},
\end{equation}
where $\Omega_{0}$ is the present-time density parameter, $h$ is the Hubble parameter and $z$ is the redshift parameter. The collision time is $\tau \simeq \frac{1}{n_{\gamma} \sigma_{T}}$, where $\sigma_{T}$ Thomson coefficient $ = \frac{e^{4}}{6 \pi m_{e}^{2}} \simeq 6.6 \times 10^{-25} cm^{2}$, and $n_{\gamma} = 4.2 \times 10^{2} cm^{-3} (1 + z)^{3}$. Substituting all this in the expression  and using $\Omega_{0}=1$ and $h=0.6$, we obtain the electrical conductivity as;
\begin{equation}
	\sigma = \frac{n_{e} e^{2}}{m_{e} n_{\gamma} \sigma_{T} } \simeq 0.33 ~ {cm}^{-1}.
\end{equation} 
The length and width of the cosmic string wake determine the length of the reconnection region. We have already mentioned that they can be calculated as long as the deficit angle is known. This deficit angle is taken to be $10^{-7}$ from the constraints obtained from the Planck data \cite{Planck_2016}.  At $z = 6.7$, this turns out to be $l = 1.4 \times 10^{10}$ cm. Initially, we do  our calculations at a fixed redshift, later we check the model by varying the redshift parameter.

\section{Multiple collisions and their temporal structure}	 \label{sec:multiplecollisions}

\begin{figure}[h]
	\centering
	\includegraphics[width=0.75\textwidth]{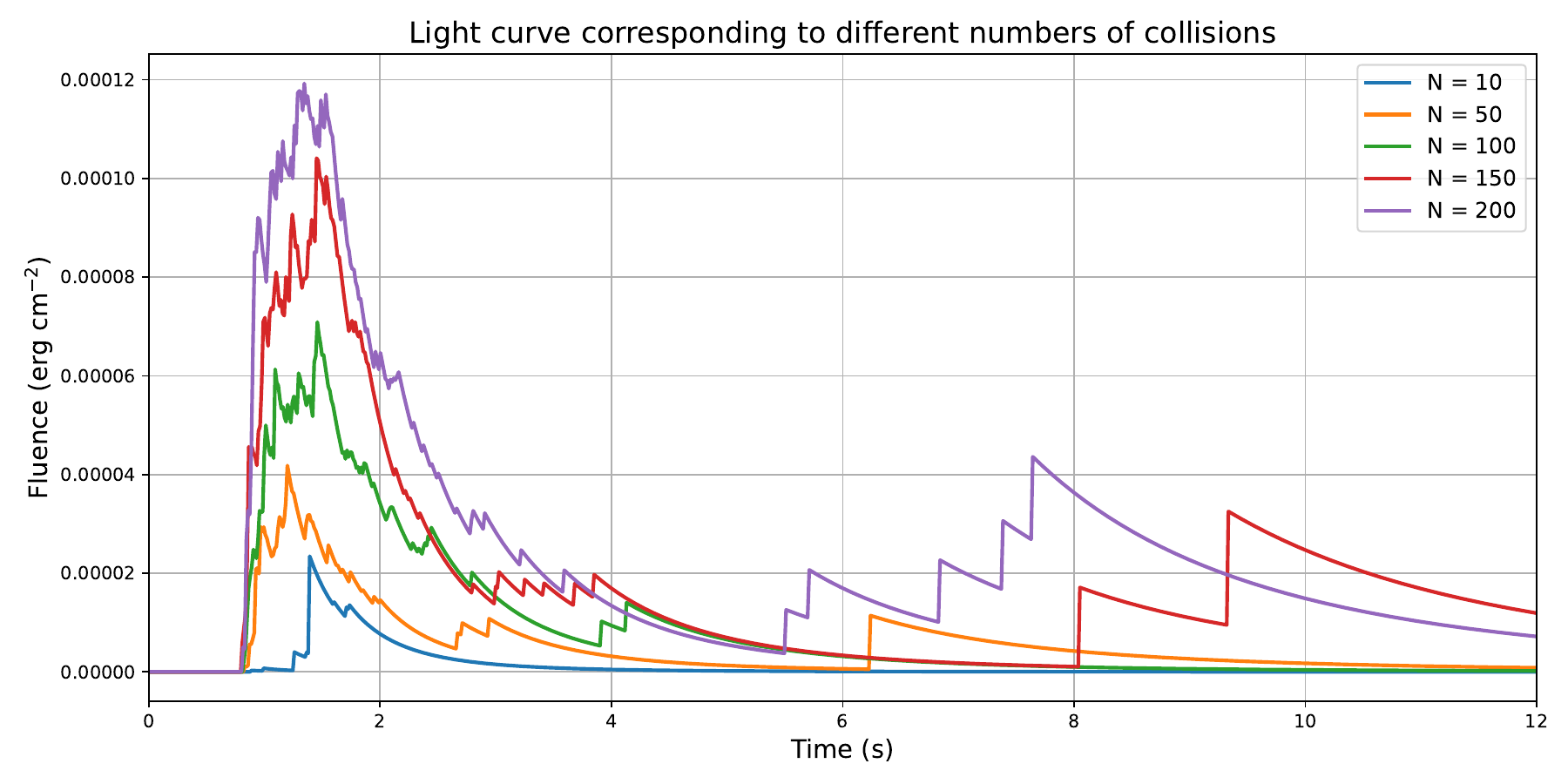}
	\caption{Energy fluence with time as multiple shocks collide and merge, followed by their decay. The shocks have random velocities but generally a faster shock collides with a slower shock in this figure. The number of random collisions is mentioned in the adjoining legends.}
	\label{fig:multipulse}
\end{figure}

The  important point that we would like to make is that the magnetic reconnection in the cosmic string wake invariably results in multiple points of magnetic reconnection. This has been illustrated in Fig.[\ref{fig:reconnection}]. So the case of the two shock collision is the simplest possible case. Multiple shocks will be generated and they may be generated from multiple points in the cosmic string wake, so there is the possibility of multiple shock collision in the cosmic string wake. However, it is important to point out that these shocks will have different velocities and therefore different $\gamma$ factors. They may not collide in the same direction. To model this kind of collisions is rather difficult. We have extended our model to include at least a few of these conditions. Extending our two shock model, we have done a simulation where multiple shocks are generated with random values of $\gamma$. These are then made to collide similar to the two shock model that we had before. We can vary the number of shocks colliding one after the other. Though not perfect, we have obtained the synthetic light curve for the multiple collisions also, as shown in Fig.[\ref{fig:multipulse}]. 
In Fig.[\ref{fig:multipulse}], we have shown the light curve for different  collision rates. Here, we have assumed that there is no correlation between the multiple shocks and each shock collides only once with the shock closest to it. We have considered all the collisions between a slow shock and a faster shock. As is expected, a larger number of collisions between the shocks only increase the total energy. We also notice that it does not change the order of the total energy generated. So it appears that the number of collisions leading to the GRB will not change the GRBs significantly. 

\begin{figure}[h]
	\centering
	\includegraphics[width=0.75\textwidth]{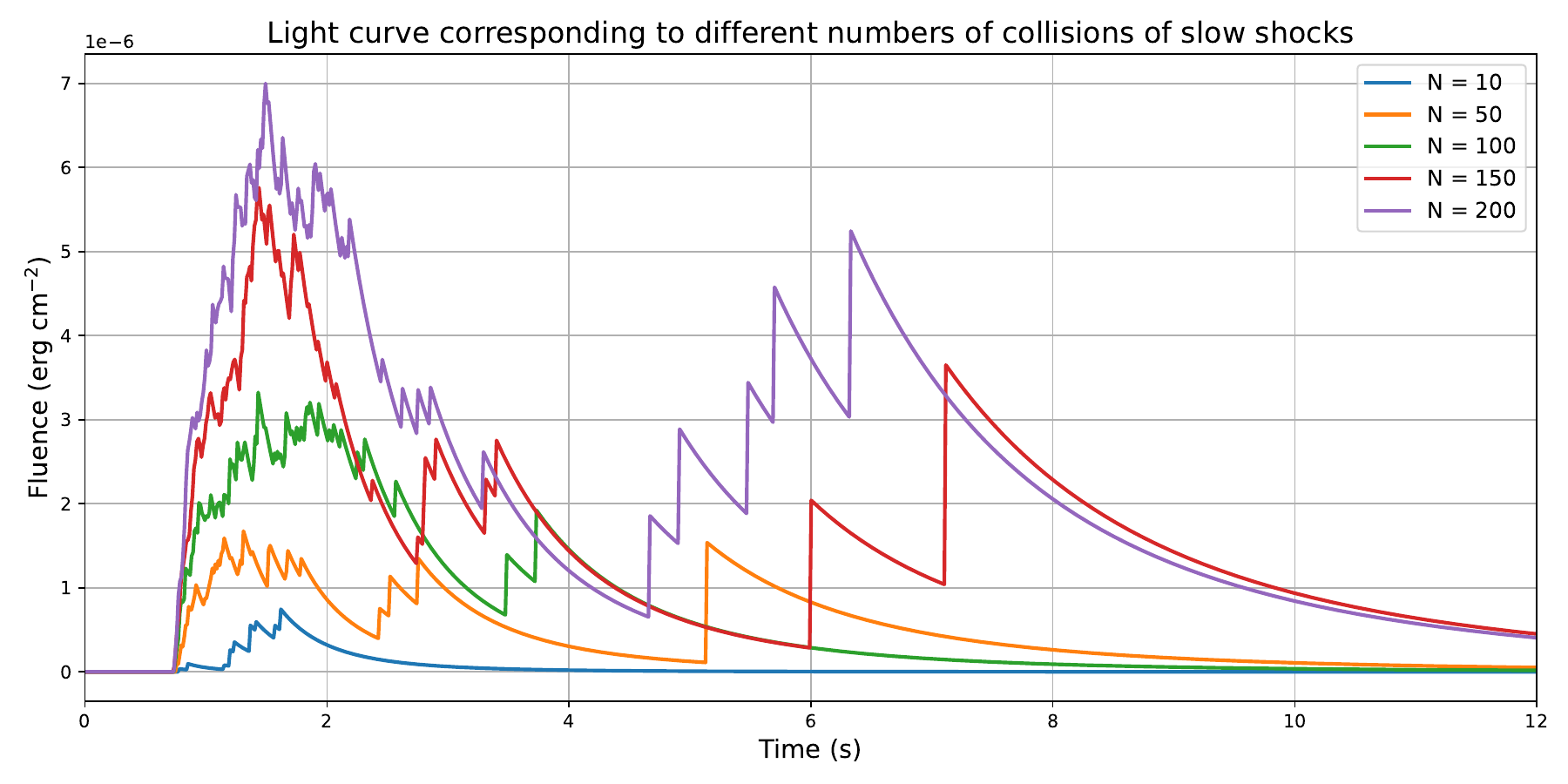}
	\caption{Energy fluence with time as multiple slow shocks collide and merge, followed by their decay. Here, we have kept the shock velocities random but always less than $0.5 $c.  The number of random collisions is mentioned in the adjoining legends}
	\label{fig:multipulse_slow}
\end{figure}

\begin{figure}[h]
	\centering
	\includegraphics[width=0.75\textwidth]{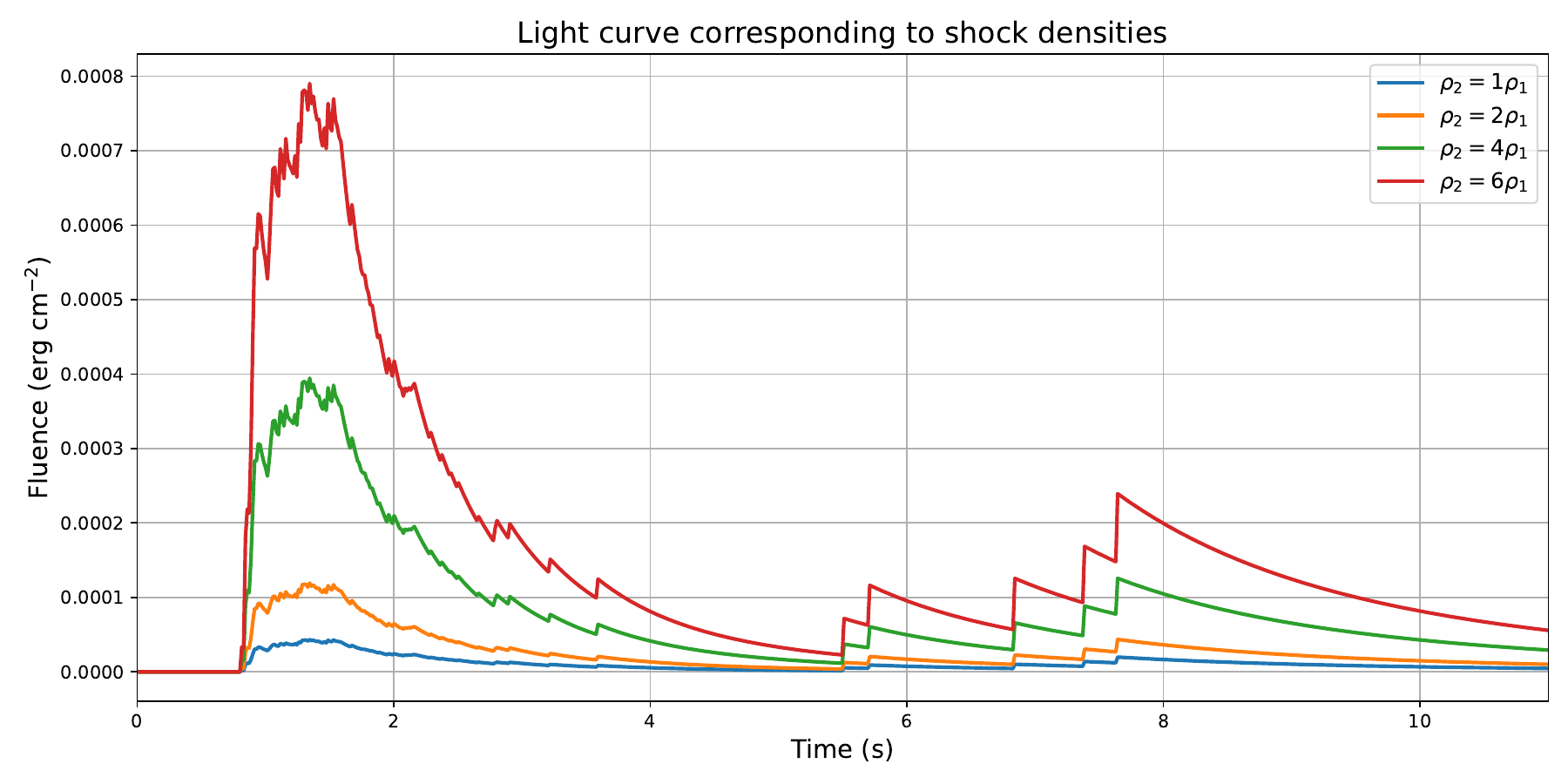}
	\caption{Energy fluence with time as multiple shocks of different density ratio collide and merge, followed by their decay. The density ratio is mentioned in the adjoining legends}
	\label{fig:multipulse_diffdensities}
\end{figure}

\begin{figure}[h]
	\centering
	\includegraphics[width=0.75\textwidth]{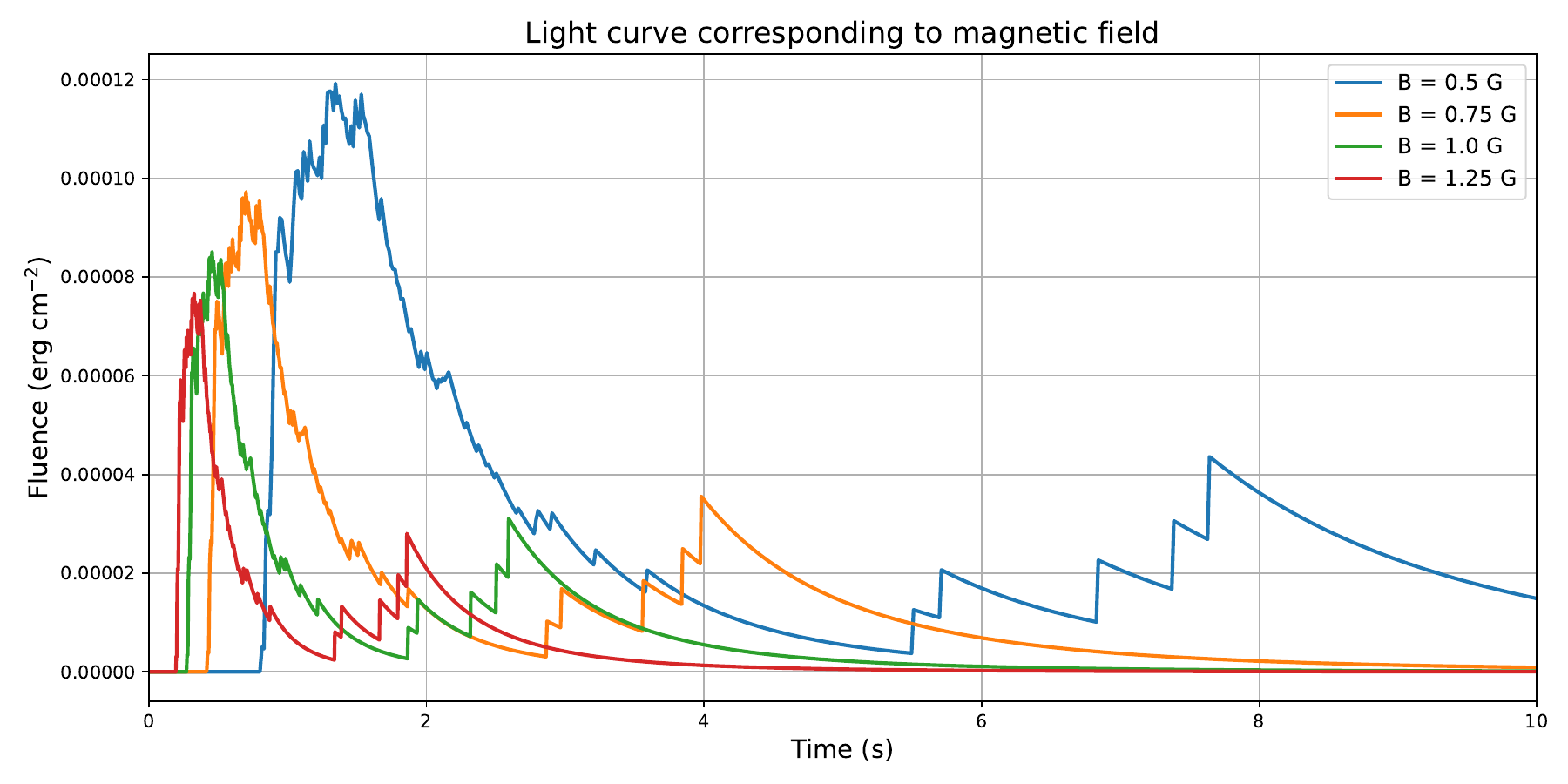}
	\caption{Energy fluence with time as multiple shocks collide and merge for different background magnetic fields. The corresponding background magnetic field  is mentioned in the adjoining legends}
	\label{fig:multipulse_diffmagfield}
\end{figure}

We also obtain similar curves with multiple slow moving shocks. By slow moving shocks we mean that the $\gamma$ factor is less than $0.5$c. These are depicted in Fig.[\ref{fig:multipulse_slow}]. Here too, we get similar light curves, only now the increasing energy is less steep than in the case where one of the shocks was a fast one.     
As mentioned before, the background densities of these shock waves have been taken as integral multiples of each other. This was for the ease of obtaining the resultant $\gamma$ factor of the colliding shocks. As a check, we also did the shock-shock collision for different densities of the shock waves. This is then shown in Fig.[\ref{fig:multipulse_diffdensities}] with the legends denoting the ratio of the densities of the shocks. As we see in this figure, we are getting different curves for the different ratios with the higher density ratios having higher energy fluence. 
Finally, we have also checked the effect of varying the background magnetic field. Again this does not change the pattern but changes the overall magnitude and timescale of the GRB. The timescale of the GRB is affected as the emergence of the shocks depends on the magnetic reconnection occurring in the cosmic string wake. For our fast reconnection model, 
$\Delta t = \frac{8}{\pi} \frac{l_r log(S)}{V_A}$. Here, $S$ is the Lundquist number and $V_A$ depends on the magnitude of the magnetic field and plasma density. This is shown in Fig.[\ref{fig:multipulse_diffmagfield}].

\begin{figure}[h]
	\centering
	\includegraphics[width=0.75\textwidth]{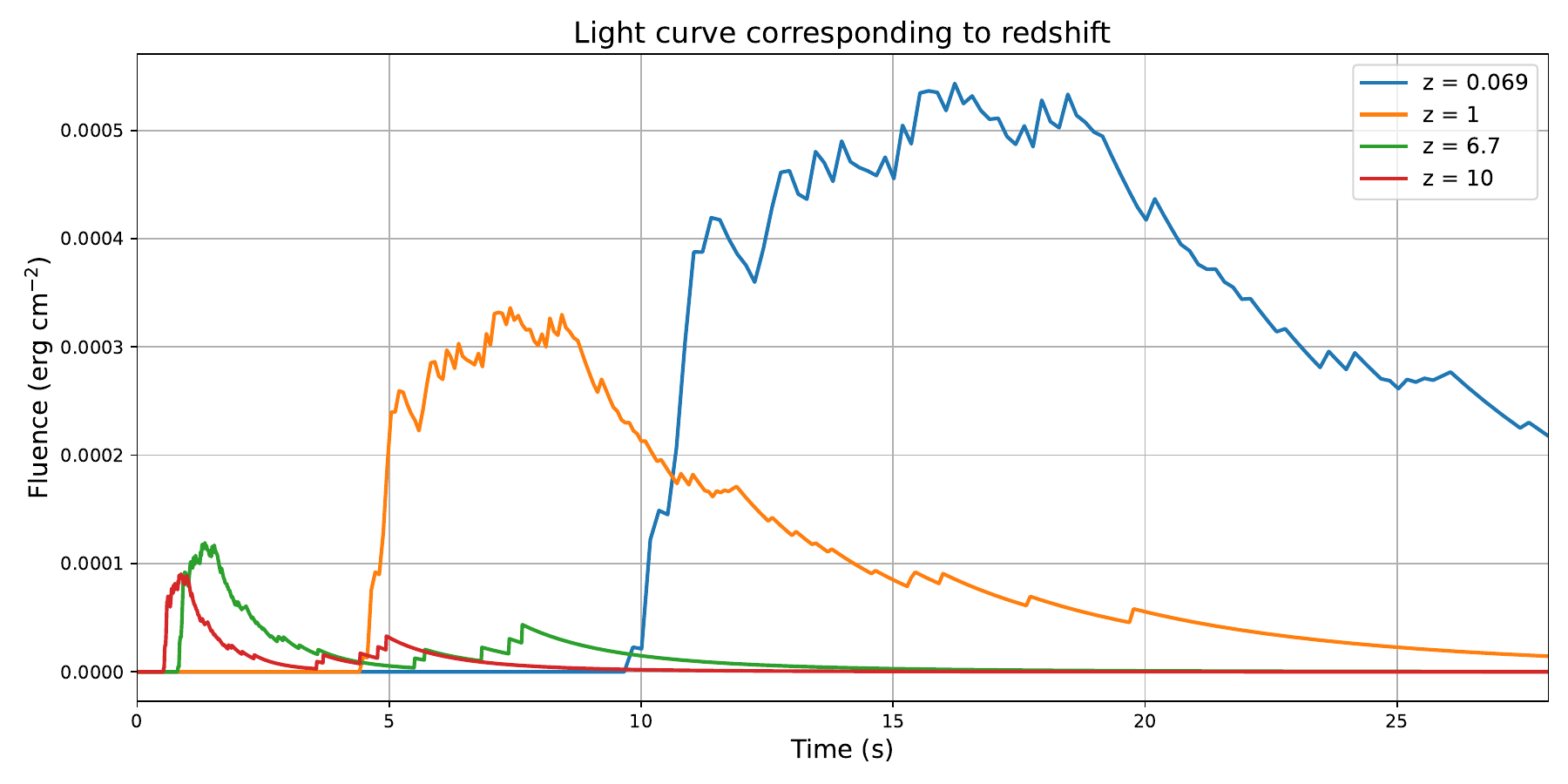}
	\caption{Energy fluence with time as multiple shocks collide and merge for different redshift values. The corresponding values of redshift $z$ are mentioned in the adjoining legends}
	\label{fig:multipulse_diffz}
\end{figure}

A general pattern is observed for the synthetic light curve obtained from the shock-shock collision in the magnetized wakes of cosmic strings. The timescale of the collision depends upon various factors. One of the most important factor being the density of the plasma. Since we want to do the calculations for a GRB that is observable, we have chosen redshift $z=6.7$. However GRB's are mostly observed at much lower $z$ values. At different redshift $z$ values, the magnetic reconnection length, the density of the plasma as well as the magnetic field may be different. Keeping the magnetic field value to be the same, we have obtained some graphs for corresponding redshift values of $z$. This is shown in Fig.[\ref{fig:multipulse_diffz}] for multiple collisions.    
Since the magnetic reconnection length changes the total energy density and the collision timescales we see that we get higher intensities for lower redshift values.

\section{Observational possibilities of the light curves}	 \label{sec:lightcurves}
 
We have theoretically modeled a single pulse of energy that can be generated from magnetic reconnection in cosmic string wakes. We have then done a superposition of such pulses for random relativistic velocities of merger to obtain a synthetic light curve for a GRB. We would like to highlight that the burst duration and the energy released from the merging shocks depends primarily on the length of the magnetic reconnection region. This in turn depends on the magnetic field lengthscale. Since, the magnetic field is generated in the wake of the cosmic string, the magnetic lengthscale would depend on the length of the wake. As mentioned previously, the magnetic fields are along the $Z-$direction. In our previous work \cite{Dilip_2023}, we have mentioned that cosmic string wakes are essentially planar wakes and so we are looking at a quasi-two dimensional process. The two dimensions considered here are the $X-Y$ plane. The hot electron current is in opposite directions in the two dimensional plane, so basically for the quasi-two dimensional plane we have approximated $\vec{B} ~=~ -\hat{i} \frac{\partial N_e}{\partial z} \frac{\partial T}{\partial y} $ to map it on the two dimensional plane [For detail study check ref.\cite{Dilip_2023}]. The magnetic lines are squeezed in the $Y-$direction due to the flow of the plasma around the cosmic string. The velocity perturbation in the $Y-$direction comes from the conical spacetime of the cosmic string. This space time has a deficit angle which is proportional to $G\mu$. Here, the $\mu$ is the mass per unit length of the string depends on the symmetry breaking scale which generated the cosmic string in the first place. We would now like to check whether any cosmic string model would indeed generate energies in the ranges where GRB's have been observed.   

As we have mentioned before, we have considered the parameters of an Abelian Higgs cosmic string. All the constraints on the cosmic string parameters are based on the results from Planck \cite{Planck_2016}. We have calculated the reconnection length from the parameters of the wake of the cosmic string. For different values of $z$, this turns out to be different. Though our simulations are based on the energy density, one can relate the energy density to the fluence associated with a GRB. The fluence associated with these light curves  can be found out by multiplying with the length associated with the wakes which are of the order of $10^{10}$ cm or less, so the fluences are found to be in the range of $\sim 10^{-5}$ erg cm$^{-2}$ or less. For redshift $z = 6.7$, the timescale is found out to be $\Delta t = 1.61$ sec. The collision time is therefore $\Delta t < 2$ sec for different Lorentz factors corresponding to the two outgoing shock waves. Thus, the obtained result may be for short-duration GRBs. One of the important factors in GRB's is the electromagnetic emission band predicted by the model. The crucial part of our work is the energy generated in the reconnection region as that is the energy which produces the shocks. As long as this energy is high enough to produce a GRB, the shocks emanating from these points will have this energy as a minimum. So we calculate the total energy generated in the string wake due to the magnetic reconnection. We base our calculation on the Petschek model but emphasis that there are  other models which can generate even faster reconnections. The width of the reconnection region ($\delta$) is given by \cite{Petschek}
\begin{equation}
	\delta = \frac{c^{2}}{4 \pi \sigma V_A M_{0}}, 
\end{equation}
where, $M_{0}$ is calculated from the diffusion region as,
\begin{equation}
	M_{0}^{2} = \frac{c^{2}}{8 \pi \sigma V_A y^{*}}, 
\end{equation}
where, $y^{*}$ is the length of reconnection region, $V_A = \frac{B}{\sqrt{4 \pi \rho}}$. We consider values at $z=6.7$. Here we have taken $B$ to be the equipartition magnetic field at the given redshift $\sim 1.95 \times 10^{-2} ~ eV^{2}$, $\rho$ is plasma density at that redshift $\sim 10^{-10} ~ eV^{4}$ and $\sigma$ is the conductivity of the plasma $\sim 0.33 ~cm^{-1}$ which has been obtained before. Thus, $y^{*}=l$ is length of reconnection region obtained is $\sim 1.40 \times 10^{10} ~cm$ from the cosmic string wake parameters. Hence, we obtain, $M_{0} \simeq 10^{-6}$ and substituting this in the width of the reconnection region, we obtain, $\delta \sim 10^{2} ~cm$. 
	
So we have obtained the energy density from magnetic reconnection in cosmic string wakes as, $\mathcal{E} \sim 10^{-15} ~erg cm^{-3}$. Since, $1 ~erg = 6.24 \times 10^{2} ~GeV$. Thus, $\mathcal{E} \sim 6.24 \times 10^{-4} ~eV cm^{-3}$. Now, for the total energy emitted from the reconnection region geometry, we use, $E = \mathcal{E} \times V$, where, $V$ is volume of the reconnection region. Thus, we obtain the total energy as $E \sim 10^{2} MeV$. This is the minimum energy band. As mentioned previously, magnetic fields can get amplified in cosmic string wakes which means that the actual magnetic field in the calculation should be at least $10$ times higher than the magnetic field we have considered. This will enhance the total energy generated in the magnetic reconnection region. This will in turn increase the energy of the shocks and therefore the total energy emitted.  
 
We have studied the available catalogue of GRBs to compare our energy scales with observed bursts for different redshifts. The table below shows the various high-redshift GRBs observed for short duration GRB's. This is not a exhaustive list of short duration GRB's but a token list to show the fluence and rest-frame burst duration ($T_{90}$) currently being observed by various satellites like FERMI, SWIFT mentioned in individual references for each GRBs. The rest-frame duration is calculated as the observed burst duration divided by $(1+z)$, accounting for cosmological time dilation. As we have also done the calculation for cosmic string wake in the rest frame. This correction allows for a fair comparison of GRB durations across different redshifts by removing the stretching effect due to the expansion of the universe. As is seen from the table, the energy scales and the timescales both indicate that magnetic reconnection in cosmic string wakes can be considered to be the progenitor for a GRB. 
\begin{table}[h]
	\centering
	\begin{tabular}{|c|c|c|c|c|c|}
		\hline
		\textbf{S.No.} & \textbf{GRBs ID} & \textbf{Redshift} & \textbf{Burst Duration ($ T_{90}$)} & \textbf{Observed Fluence}	& Ref. \\
		&  	&	& \textbf{(sec.)} & \textbf{($\times 10^{-6}$ erg cm$^{-2}$)}	&  \\ \hline
		1.  & 090429B    & 9.4   & 0.53 $\pm$ 0.09 & 0.31 $\pm$ 0.03	& \cite{Cucchiara_2011} \\ \hline
		2.	& 090423     & 8.1   & 1.13 $\pm$ 0.12 & 0.59 $\pm$ 0.04	& \cite{Salvaterra_2009} \\ \hline
		3.	& 080913     & 6.7   & 1.04 $\pm$ 0.13 & 0.56 $\pm$ 0.06	& \cite{Greiner_2009} \\ \hline
		4.	& 220521A    & 5.6   & 1.81 $\pm$ 0.07 & 0.98 $\pm$ 0.12	& \cite{Minaev_2022} \\ \hline
		5.	& 240419A    & 5.17  & 0.48 $\pm$ 0.22 & 0.08 $\pm$ 0.02	& \cite{Parsotan_2024} \\ \hline
		6.	& 090205     & 4.65  & 1.55 $\pm$ 0.31 & 0.19 $\pm$ 0.03	& \cite{DAvanzo_2010} \\ \hline
	\end{tabular}
\end{table}

\begin{figure}[h]
	\centering
	\includegraphics[width=0.75\textwidth]{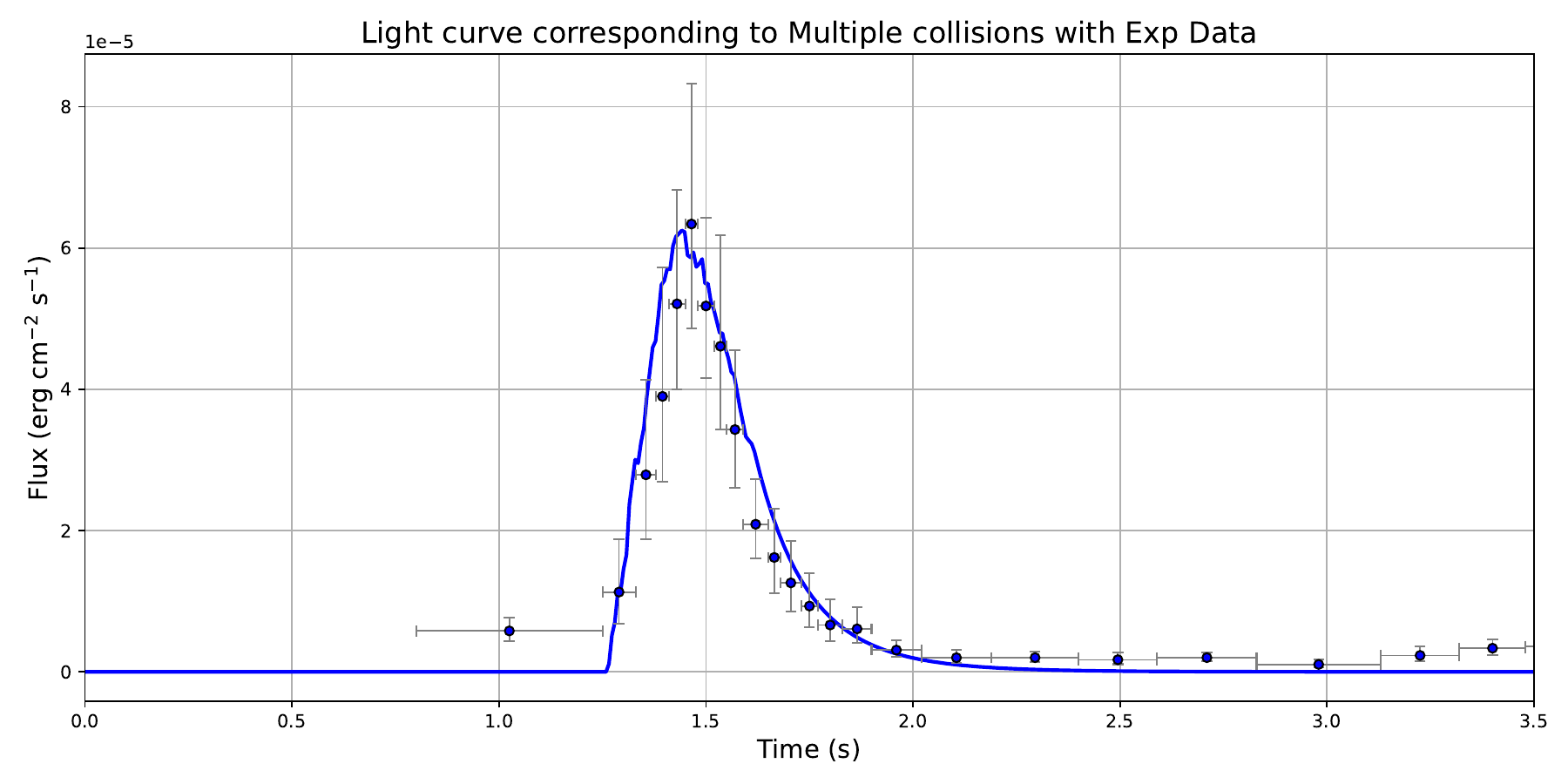}
	\caption{The graph shows the best fit for the data from GRB 081215 to our model. The $\alpha$ parameter is quite high with the current fit being given for $\alpha = 13$ and $\beta = 2$. }
	\label{fig:data_fit}
\end{figure}

We have also looked at the published data and made an attempt to fit the data from our model. The model does not incorporate many of the finer points of the plasma dynamics that occur in a cosmic string wake but even then it can fit the data to a great extent. We have calculated the flux and matched it with the data from the reference \cite{chang}. This is shown in Fig.\ref{fig:data_fit}. We have fitted the data from the GRB 081215. As is seen in the figure, we do not get a perfect match. For data fitting, we have mostly varied the shock collision velocities and the magnetic field as these are the unknown parameters of our model. An interesting observation was that a constant magnetic field does not fit the data well. This is not surprising as the magnetic field in the cosmic string wake will not necessarily be a constant field. 
The collision timescale and length of the reconnection region remain fixed by the parameters of the cosmic string. But due to diffusion processes, the magnetic field does not remain the same as the shocks travel in an outward direction. We have varied the free parameters $\alpha$ and $\beta$ to
fit the data from the GRB.  This fit is shown in fig. \ref{fig:data_fit}. The fit that we have obtained is for a value of $\alpha = 13$ and $\beta = 2 $. Thus the fit strongly depends on the way
the magnetic field decays. We have only fitted the dominant pulse. A best fit for all the data will require including other dissipative forces.  

Most of our detailed calculations are done for the redshift value of $z = 6.7$.  Recently some studies on cosmic strings have been done for lower values of $z$ with the limitation of $z > 15$ \cite{Oscar_2021}. For the GRB emission, the most important change due to the redshift would be the value of the background plasma density. We have found that changes in the plasma density  only changes the maximum energy associated with the light pulse. The mechanism for the generation of the GRB remains the same irrespective of the redshift. In a future work, we plan to improve the model by 
incorporating other features of the plasma. This will enable us to fit the  
experimental data with greater accuracy. We believe that in addition to the other signatures of 
a cosmic string, this will help us in determining the presence of a cosmic string in the current universe.

\section{Summary and Conclusions} 					\label{sec:conclusion}

To summarize, we have worked out a  model for the generation of a light pulse from the collision of shocks in the magnetized wake of a cosmic string at different redshifts. Gamma-ray bursts (GRBs) arising from cosmic strings have been  studied in the literature in the context of superconducting cosmic strings before. However in our model we consider non-superconducting cosmic strings where the magnetic fields arise due to the Biermann battery mechanism. This will lead to one significant difference. The GRB's generated from the cusps of super conducting strings are strongly collimated in a narrow region. The GRB's generated in the wake region will not be so constrained. Though the energy would appear to be collimated but the opening angle would be larger than in the case of the superconducting cosmic strings. In the current work, we have not calculated the event rate of the GRB's from the cosmic string wake. This is because unlike the superconducting cosmic string case, there are many more parameters that need to be evaluated precisely before an event rate can be calculated. Primary amongst this is the understanding of the exact kind of magnetic reconnection that is happening in the cosmic string wake. This study itself will require large scale MHD simulations which are currently beyond the scope of this work. We plan to follow it up later and we will try to include all the possible parameters in the collision model. There are other ways in which we can also pinpoint whether the GRB is from a cosmic string wake. As mentioned in the Introduction section, the wakes will leave an imprint on the 21cm redshift survey maps. Mapping these imprints with GRB bursts may lead us to a cosmic string wake. Since any cosmic string has 
a wake structure behind it, we do not need to look at Superconducting strings only. This we
believe increases the chances of identifing cosmic strings from the current observational
data. We will now conclude with a brief summary of the work.   

Magnetic field generated in cosmic string wakes are oppositely directed on both sides of the wakes. As the string moves through the plasma, the magnetic field lines are pushed together and magnetic reconnection can occur. This reconnection results in the generation of a large amount of kinetic energy. Different models of magnetic reconnection gives rise to multiple shocks. We show that the collision of these shocks can result in the generation of a strong pulse of energy. Multiple shocks colliding may give rise to a GRB light curve. Using the various parameters of shock waves due to magnetic reconnection in cosmic string wakes we have modeled the light pulse and the synthetic light curves. We find a generic pattern for the GRB light curve. It depends primarily on the density of the background plasma and the magnetic field. There is a weak dependence on the ratio of the densities of the colliding shocks, similarly there is a weak dependency on the velocities of the colliding shocks. We have fitted the data of an observed GRB to our model. Though the fit is not perfect, it shows that the generic pattern of the GRB is reflected in the data too. This indicates 
that the model can be improved to fit the data from existing GRB's by incorporating further details 
of the plasma dynamics.  

Since this is the first work on the GRB's from non-superconducting cosmic string we have considered some basic assumptions for the current work. We have only considered the collision of shocks in one direction and have assumed that there is no correlation in the multiple shock mergers. This means that multiple collision will just enhance the overall energy of the outburst. A more detailed simulation would include various other factors. It is possible that after the initial collision between two shocks, the  merged shock will coast for some time before colliding with another shock generated from a different point. This will require changes in the timescales and will therefore change the light curve considerably. As we have mentioned in the text, the cosmic string wake is bounded by strong shocks on both sides. This will lead to cases of shock reflections and the collision with a reflected shock. The shocks are also oblique in nature so it may not be a smooth merger. We have not considered any of these cases. We have not considered several other cases such as Mach stems, all of which can occur during shock collisions. We believe that detailed studies are necessary to improve the lightcurve further.  Since we have found that the calculated fluence and the burst duration are well within the ambit of GRB's observed by various experimental groups, we would like to incorporate some more features of shock collisions in these simulations in a later work. Our primary objective in this work was to show conclusively that magnetic reconnection in the wakes of cosmic strings can indeed be the progenitor for a GRB.

\textbf{Acknowledgment}: 
We acknowledge funding from the DST-SERB/ANRF Power Grant no.SPG/2021/002228, D.K is supported by the Power Grant no. SPG/2021.002228, of the Government of India.

%
%


\begin{thebibliography}{100}
%
\bibitem{cosmicstrings} A. Vilenkin, and E. Paul S. Shellard. Cosmic strings and other topological defects. Cambridge University Press, 2000.

\bibitem{lensing} M. V. Sazhin, O. S. Khovanskaya, M. Capaccioli, G. Longo, M. Paolillo, G. Covone, N. A. Grogin, E. J. Schreier, Monthly Notices of the Royal Astronomical Society 376, 4, 1731–1739 (2007).

\bibitem{wakes} R. H. Brandenberger, J. Silk, and N. Turok, Astrophys. J. 322, 1 (1987); T. Vachaspati, Phys. Rev. D 45, 3487 (1992).

\bibitem{branden1} R. H. Brandenberger, R. J. Danos, O. F. Hernández, and G. P. Holder, J.Cosmol. Astropart. Phys. 12 (2010) 028.

\bibitem{dacunha} da~Cunha, D. C.~N. 2020, Journal of Cosmology and Astroparticle Physics, 2020, 016.

\bibitem{Planck_2016} Planck Collaboration, Ade, P. A. R., Aghanim, N., et~al. 2016, A\&A, 594, A13.

\bibitem{Zhou_2023} L. Zhou, D. Lin, X. Yang, G-Y. Li, K. Liu , J. Li and E-W. Liang, The Astrophysical Journal 957, 109 (2023).  

\bibitem{Sazhina_2014}Sazhina, O.S., Scognamiglio, D. Sazhin, M.V. Eur. Phys. J. C 74, 2972 (2014).
\bibitem{Blasi_2021} Blasi, Simone, Vedran Brdar, and Kai Schmitz. Physical Review Letters 126.4 (2021): 041305.

\bibitem{vachaspati} T. Vachaspati, Phys. Rev. Lett. 101, 141301 (2008). 

\bibitem{Harrison}  E. R. Harrison, Monthly Notices of the Royal Astronomical Society 165, 185 (1973).

\bibitem{Sovan_2020} Sau, S., and Sanyal, S. 2020, The European Physical Journal C, 80. 

\bibitem{ohira} Y. Ohira, The Astrophysical Journal 911, 1, 26 (2021).

\bibitem{Dilip_2023} Kumar, D., and Sanyal, S. 2023, The Astrophysical Journal, 944, 183.

\bibitem{Soumen_2022} S. Nayak, S. Sau and S. Sanyal, Astroparticle Physics 146, 102805 (2023).

\bibitem{Paczynski_1998} Paczyński, B. The Astrophysical Journal 494, L45 (1998). 

\bibitem{Babul_1987} Babul, A., Paczynski, B., \& Spergel, D., The Astrophysical Journal Letters, 316, L49 (1987).

\bibitem{berezinsky} V. Berezinsky, B. Hnatyk and A. Vilenkin, Phys. Rev. D 64, 043004, (2001).

\bibitem{Granot} J. Granot, The Astrophysical Journal Letters, 816, L20 (2016).

\bibitem{yamada_review} Yamada, M., Russell K., and Hantao Ji.  Reviews of modern physics 82, 1, 603-664 (2010).

\bibitem{Petschek} H. E. Petschek, On the Physics of Solar Flares, In Proc. of AAS-NASA Symp. (Vol. 425). NASA Spec. Pub. (1964). 

\bibitem{piran} Piran, T. Reviews of modern physics 76, 4, 1143-1210 (2004).

\bibitem{vilenkin} A. Vilenkin. Physics Reports, 121, 263-315, (1985).

\bibitem{ssanyal1} B. Layek, Soma Sanyal, A. M Srivastava, Physical Review D 63, 8, 083512 (2001).

\bibitem{ssanyal2} B. Layek, Soma Sanyal, A. M Srivastava, Physical Review D 67, 8, 083508 (2003).

\bibitem{sornborger} Sornborger, A., Brandenberger, R., Fryxell, B., and Olson, K., The Astrophysical Journal 482, 1, 22-32 (1997).

\bibitem{Parker_1957} Parker, E.~N., Journal of Geophysical Research, 62, 509 (1957).

\bibitem{che} H. Che, Phys. Plasmas 24, 082115 (2017).

\bibitem{lundquistno} Bhattacharjee, A., Huang, Yi-Min, Yang, H., and Rogers, B., Physics of Plasmas 16, 11, 112102 (2009).

\bibitem{Bist2025} D. Bisht, D. Kumar, S. Nayak, and S. Sanyal, International Journal of Modern Physics D (in press) https://doi.org/10.1142/S021827182550097X .

\bibitem{Eyink} G.L. Eyink, A. Lazarian, E.T. Vishniac, The Astrophysical Journal 743 (1), 51 (2011).

\bibitem{Workman} J. C. Workman, E. G. Blackman, and C. Ren, PHYSICS OF PLASMAS 18, 092902 (2011).

\bibitem{Hartigan} P. Hartigan et al  ApJ 823 148 (2016).

\bibitem{Rahaman} Rahaman, Sk M., Jonathan G., and Paz B., Monthly Notices of the Royal Astronomical Society 528, 1, 160–179 (2024).

\bibitem{meng}Ying Meng, Jun Lin and Feng Yuan, Res. Astron. Astrophys. 15 207 (2015).

\bibitem{yuan}Yuan, F.,  Zhang, B. 2012, ApJ, 757, 56.

\bibitem{Asaf} Asaf Pe’er, Killian Long, and P. Casella, The Astrophysical Journal 846, 54, (2017).

\bibitem{Kulsrud} Kulsrud R. M 2001, Earth Planets Space, 53, 417–422, 200. 

\bibitem{Lizarraga_2016} Lizarraga, J., Urrestilla, J., Daverio, D., Hindmarsh, M., \& Kunz, M. 2016, Journal of Cosmology and Astroparticle Physics, 2016, 042. 

\bibitem{beniamani} P. Beniamini and J. Granot, Monthly Notices of the Royal Astronomical Society 459 (4), 3635-3658 (2016). 

\bibitem{vieu} T. Vieu, S. Gabici and V. Tatischeff, Monthly Notices of the Royal Astronomical Society 494, 3166–3176 (2020).

\bibitem{kobayashi}S. Kobayashi, T. Piran and R. Sari, The Astrophysical Journal, 490, 92-98 (1997).

\bibitem{Grasso_2001}Grasso, D.,  Rubinstein, H. R. 2001, Physics Reports, 348, 163-266.

\bibitem{Oscar_2021} Oscar F. Hernandez, Monthly Notices of the Royal Astronomical Society, Volume 508, Issue 1, (2021). 

\bibitem{Cucchiara_2011} Cucchiara, A., Levan, A. J., Fox, D. B., et al. 2011, The Astrophysical Journal, 736, 7. 

\bibitem{Salvaterra_2009} Salvaterra, R., Della Valle, M., Campana, S., et al. 2009, Nature, 461, 1258.

\bibitem{Greiner_2009} Greiner, J., Krühler, T., Fynbo, J. P. U., et al. 2009, The Astrophysical Journal, 693, 1610.

\bibitem{Minaev_2022} Minaev, P., Pozanenko, A., Mozgunov, G., et al.\ 2022, GRB Coordinates Network, Circular Service, No. 32105, GRB 220521A. 

\bibitem{Parsotan_2024} Parsotan, T., Barthelmy, S.~D., Krimm, H.~A., et al.\ 2024, GRB Coordinates Network, Circular Service, No. 36255, GRB 240419A: Swift-BAT refined analysis, 36255, 1.

\bibitem{DAvanzo_2010} D’Avanzo, P., Perri, M., Fugazza, D., et al. 2010, A\&A,
522, A20.

\bibitem{chang}X-Z Chang, H-J Lu, X. Yang, J-M Chen, and E-W Liang, The Astrophysical Journal Supplement Series 275:9 (2024).

\end{thebibliography}
\end{document}